\documentclass{article}
\pdfoutput=1
\usepackage[a4paper, total={6.5in, 9in}]{geometry}
\usepackage[utf8]{inputenc}
\usepackage{graphicx}
\usepackage{amsmath}
\usepackage[affil-it]{authblk}

\title{Designing Silicon Photonic Devices using Artificial Neural Networks}
\author[1]{Alec M. Hammond}
\author[1,*]{Ryan M. Camacho}
\affil[1]{Department of Electrical \& Computer Engineering, Brigham Young University, Provo, Utah}
\affil[*]{Corresponding author: camacho@byu.edu}
\date{}

\begin{document}

\maketitle

% ------------------------------------------------------- %
% Abstract
% ------------------------------------------------------- %

\begin{abstract}
	
We develop and experimentally validate a novel artificial neural network (ANN) design framework for silicon photonics devices that is both practical and intuitive. As case studies, we train ANNs to model both strip waveguides and chirped Bragg gratings using a small number of simple input and output parameters relevant to designers. Once trained, the ANNs decrease the computational cost relative to traditional design methodologies by more than 4 orders of magnitude. To illustrate the power of our new design paradigm, we develop and demonstrate both forward and inverse design tools enabled by the ANN. We use these tools to design and fabricate several integrated Bragg grating devices. The ANN's predictions match very well with the experimental measurements and do not require any post-fabrication training adjustments.

\end{abstract}

% ------------------------------------------------------- %
% Introduction - background and why is this important
% ------------------------------------------------------- %

\section{Introduction}

Silicon photonics has become a viable technology for integrating a large number of optical components in a chip-scale format.  Driven primarily by telecommunications applications, a growing number of CMOS fabrication facilities dedicated to silicon photonics are now in operation and available for researchers and engineers to submit photonic integrated circuit (PIC) designs to be fabricated \cite{chrostowski_silicon_2015}.  

Designing the silicon photonics components and circuits, however, remains a major bottleneck.  Current design flows are complicated by computational tractability and the need for designers with extensive experience \cite{bogaerts_silicon_2018}. Unlike their electronic counterparts, photonic integrated circuits require computationally expensive simulation routines to accurately predict their optical response functions.  The typical time to design integrated photonic devices now often exceeds the time to manufacture and test them.

To address this challenge, we propose and experimentally validate a new design paradigm for silicon photonics that leverages artificial neural networks (ANN) in an intuitive way and is at least four orders of magnitude faster than traditional simulation methods.   Our design paradigm only requires a small number of input \emph{and} output neurons corresponding to descriptive parameters relevant to the designer.

This new approach provides benefits such as rapid prototyping, inverse design, and efficient optimization,  but requires a more sophisticated ANN than previous approaches.  We demonstrate practical confidence in our method's accuracy by fabricating and measuring devices that experimentally validate the ANN's predictions.  As illustrative examples, we design, fabricate, and test chirped integrated Bragg gratings. Their large parameter spaces and nonlinear responses are typical for devices that are computationally prohibitive using other techniques.  The experimental results show remarkable agreement with the ANN's predictions, and to the authors' knowledge represent the first experimental validation of photonic devices designed using ANN's.

 This work builds on recent theoretical results showing that it is possible to model nanophotonic structures using ANNs.  For example, Ferreira et al. \cite{da_silva_ferreira_towards_2018} and Tahersima et al. \cite{tahersima_deep_2018} demonstrated that ANNs could assist with the numerical optimization of waveguide couplers and integrated photonic splitters respectively. In both cases the input parameter space was the entire 2D array of grid points,  showing the power of ANNs in blind "black box" approach, though limiting the designer's ability to intuitively adjust input parameters. A related approach has also been applied to periodic photonic structures, which are often difficult to efficiently model \cite{inampudi_neural_2018, ferreira_computing_2018}.

Two recent theoretical papers by Zhang et al \cite{zhang_spectrum_2018}, and  Peurifoy et al.  \cite{peurifoy_nanophotonic_2018} go beyond simply optimizing over a large parameter space, and used ANNs to calculate complicated spectra using a smaller number of intuitive, smoothly varying input parameters. Even though in both cases each wavelength point in the calculated spectra required its own ANN output neuron, the usefulness of using ANNs to model systems with intuitive input parameters was demonstrated.

To illustrate the power of our new design paradigm, we demonstrate both forward and inverse design tools that use a chirped Bragg grating ANN as a computational backend. The forward design tool is interactive, and was used to design our fabricated circuits. The inverse design tool quickly constructs a temporal pulse compressing chirped Bragg grating within specified design constraints --- a task typically too computationally expensive for traditional methods.

% ------------------------------------------------------- %
% Results - what we did and what happened
% ------------------------------------------------------- %

\section{Results}

\subsection*{Overview}
To motivate our approach, we first describe a neural network that models the effective index of a silicon photonic strip waveguide with various widths and thicknesses.  While waveguide simulation is already straightforward from a designer's perspective, the model illustrates the advantages of our approach and is a key building block for more advanced ANN models described below which are less straightforward using existing techniques.  These advantages include the ANN's computational speedup of over 4 orders of magnitude, and the simplification and speedup of other complicated simulation routines that rely on effective index calculations.  A more advanced example that is computationally intractable via traditional methods is then given, in which we demonstrate an ANN that models the complex relationship between a chirped silicon photonic Bragg grating's design parameters and its corresponding spectral response. Many designers leverage silicon photonic chirped Bragg gratings to equalize optical amplifier gain \cite{rochette_gain_1999}, compensate for semiconductor laser dispersion \cite{tan_chip-scale_2008,strain_design_2010}, and enable nonlinear temporal pulse compression \cite{tan_monolithic_2010,b._j._eggleton_g._lenz_n._m._litc_optical_2000}.

\begin{figure}
\centering
\includegraphics{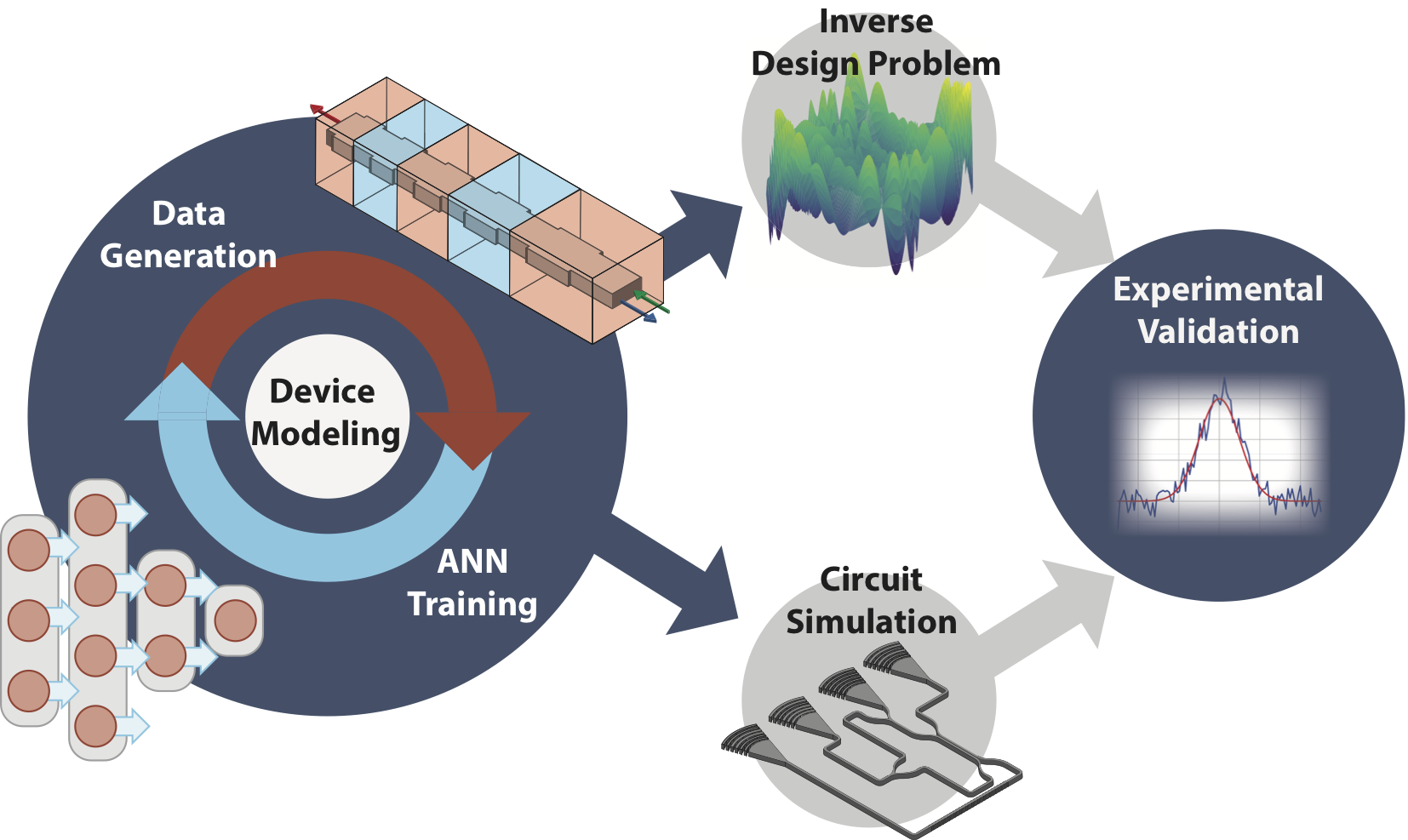}
\caption{The process overview describing the new design methodology. First, datasets are generated using traditional numerical methods (described in Methods). From this dataset, a neural network is trained to characterize the device under consideration. Figures \ref{fig:NN_neff_err} \& \ref{fig:panelFig_BG} illustrate this process for a strip waveguide and a chirped grating respectively. Often, the designer iterates between these two steps until an appropriate model is developed. Once the model is ready, several design applications, like circuit simulations and inverse design solutions, are available. The designs are then fabricated to validate the model's results. From here, the model can be shared and extended. }
\label{fig:flowChart}
\end{figure}

Figure \ref{fig:flowChart} illustrates the new design methodology. First, we iterated between generating an appropriate dataset and training the ANN until the model adequately characterized the device. Next, we used the ANN to simulate circuits and solve inverse design problems. Finally, we fabricated devices to validate the results. 

\subsection*{Waveguide Neural Network}

\begin{figure}
\centering
\includegraphics{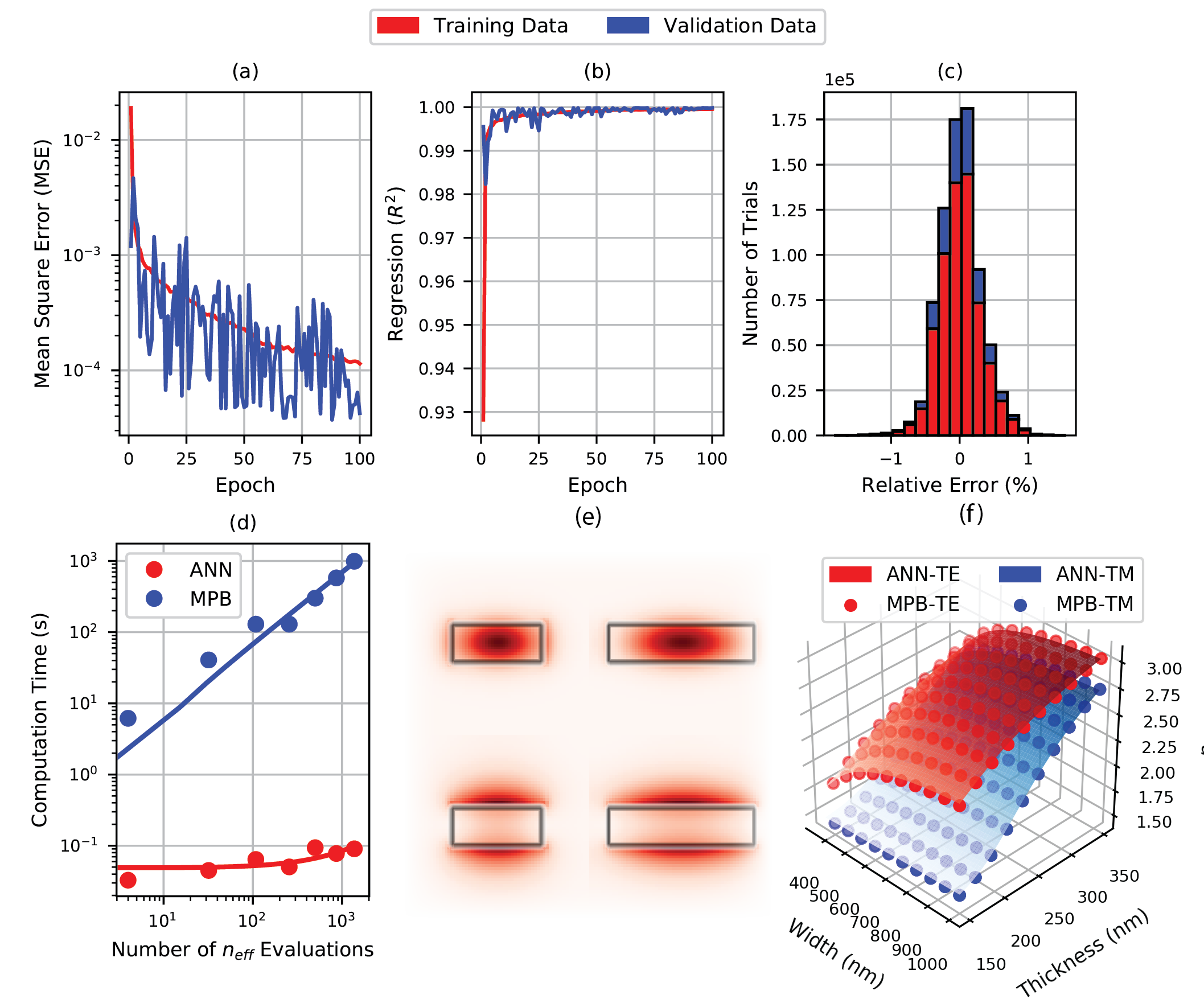}
\caption{Waveguide artificial neural network training results demonstrated by the training convergence with reference to the mean square error (a), the coefficient of determination (b), and the residual errors after training (c). Panel (d) compares the computational cost for the ANN and the eigenmode solver thas is used to simulate the mode profiles (e). Panel (f) exhibits the effective index profiles as a function of a waveguide geometry at 1550 nm for the first TE and TM modes.}
\label{fig:NN_neff_err}
\end{figure}

We first report on a simple waveguide neural network capable of estimating the effective index of any arbitrary silicon photonic waveguide geometry for a variety of modes. Specifically, we modeled the relationship between the waveguide's width, thickness, and operating wavelength and the effective index for the first two TE and TM modes. We note that including wavelength as an input parameter is a unique and enabling strategy not previously adopted (see Discussion section below). Figure \ref{fig:NN_neff_err} (f) compares the ANN's predicted effective index to a its corresponding simulation. The first TE and TM mode for any silicon photonic waveguide with a width between 350 nm and 1000 nm and a thickness between 150 nm and 350 nm are demonstrated. The network estimates a smooth response for both modes simultaneously, even for data points outside of its training set. The ANN's smooth output also produces smooth analytic derivatives, which are essential for calculating group index profiles and for gradient-based optimization routines. \par

We implemented various tests to validate the network's accuracy. First, we split the initial dataset  into a training set and a validation set. While the network evaluated both sets after each epoch (i.e. training iteration) only the training set's results were used to update the network's weights. We monitored the validation set's results to assess overfitting. To better understand the network's performance after each iteration, we recorded each epoch's mean-square-error (MSE) and coefficient of determination ($R^2$). Figure \ref{fig:NN_neff_err} (a) and (b) illustrate the MSE and $R^2$ respectively after each epoch. To prevent overfitting, we stopped training at 100 epochs, where the MSE and $R^2$ appear to converge. At this point, the network demonstrated a MSE of $1.323 \times 10^{-4}$ for the training set and $7.490 \times 10^{-5}$ for the validation set. The final $R^2$ values for the training data and validation data were $0.9996$ and $0.9997$ respectively. The MSE and $R^2$ evolution for both the training set and validation set converge well, indicating little to no overfitting. Figure \ref{fig:NN_neff_err} (c) illustrates the relative error for both the training and validation sets after the final epoch. Both the training set errors and validation set errors are similarly distributed and tightly bounded between $-1\%$ and $1\%$, once again indicating little to no overfitting.\par

With confidence in the waveguide neural network's prediction accuracy, we benchmarked its speed and found that a single neural network evaluation was $10^4$ times faster on average than the corresponding finite difference eigenmode simulation. Figure \ref{fig:NN_neff_err} (d) compares the computation speed for the ANN to the eigenmode solver, Meep Photonic Bands (MPB). This significant speedup enables many simulation techniques, like the layered dielectric media transfer matrix method (LDMTMM) \cite{helan_comparison_2006} or the eigenmode expansion method (EMM) \cite{gallagher_eigenmode_2003}, where photonic components are discretized into individual waveguides. Using the ANN, a transfer matrix for each waveguide can be quickly generated and cascaded to formulate a fairly accurate response for the device. In addition, modeling fabrication variations is now much quicker since existing Monte Carlo sampling routines can leverage the ANN's speed.\par

\subsection*{Bragg Grating Neural Network}

Modeling the relationship between a Bragg grating's physical design parameters and its corresponding responses is difficult since no one-to-one mapping exists. Consequently, many designers resort to black-box optimization routines that strategically search the parameter space for viable design options. As a result, inverse design problems --- where a simulation needs to run each iteration --- become intractable for even modest size gratings. If a full 3D FDTD simulation is performed, for example, each optimization iteration can take between 8-12 hours on typical desktop computing systems. In addition, the optimization routines tend to inefficiently simulate redundant test scenarios for different design problems. We train and demonstrate a Bragg ANN, however, that can predict a grating's response on the order of milliseconds on the same system, enabling much faster solutions to more complex design problems. We fabricate various test devices and validate our neural network's predictions. \par 

\begin{figure}
	\centering
	\includegraphics{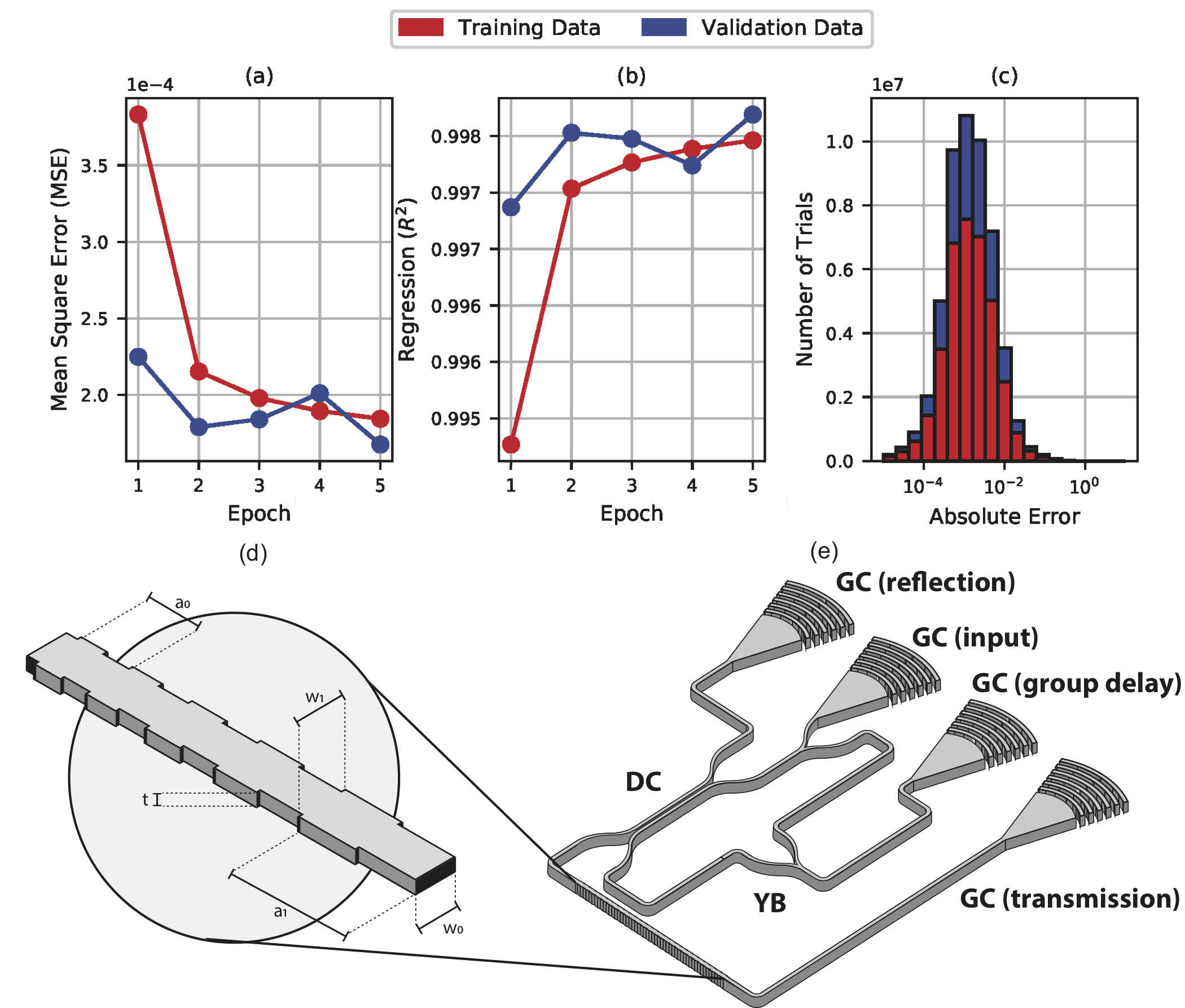}
	\caption{Bragg grating artificial neural network training results demonstrated by the training convergence with reference to the mean square error (a), the coefficient of determination (b), and the absolute error after training (c). (d) illustrates the different adjustable grating parameters and (e) illustrates the interrogation circuit used to extract the reflection, transmission, and group delay profiles simultaneously from the chirped Bragg grating. A grating coupler (GC) feeds light into various Y-branches (YB) and directional couplers (DC) such that the transmission and reflection spectra can both be extracted from the chirped Bragg grating (BG). Half of the reflected signal is sent through a Mach-Zehner Interferometer (MZI). The output of which is used to extract the group delay.  .}
	\label{fig:panelFig_BG}
\end{figure}

Using the waveguide neural network, we generated a dataset to train our Bragg grating neural network to predict the reflection spectrum and group delay response of a silicon photonic, sidewall-corrugated, linearly chirped Bragg grating, as illustrated in Figure \ref{fig:panelFig_BG} (d).  We note that generating the training dataset was approximately 2 orders of magnitude faster using the waveguide ANN reported above rather than traditional methods.   To smooth apodization dependent ringing, we pre-processed the training data. More information regarding this step is provided in the Supplementary Material. We parameterized the gratings by length of the first grating period ($a_0$), length of the last grating period ($a_1$), number of grating periods $NG$, and grating corrugation width difference( $\Delta w = w_1 - w_0$).  We designed the network to receive these four parameters along with a single wavelength point as inputs.  The network has two outputs: reflected optical power and group delay. \par

Similar to the waveguide network, we divided the dataset into a training set and validation set. We tracked both the MSE and the $R^2$ metrics after each epoch. The Bragg training set was much larger than the waveguide training set, owing to the larger parameter space. Consequently, the MSE converged within the first few epochs and we stopped training after just five epochs to prevent overfitting. The final MSE for the training and validaton sets was $1.845 \times 10^{-4}$ and $1.677\times10^{-4}$ respectively. The final $R^2$ was 0.9975 and 0.9977 respectively. Once again, the MSE and $R^2$ evolution for both the training set and validation set converge well, indicating little to no overfitting. Figure \ref{fig:panelFig_BG} (a-b) illustrates the network's MSE and $R^2$ evolution. Figure \ref{fig:panelFig_BG} (c) illustrates the absolute error for both the training sets and validation sets. We calculated the absolute error because several training samples were at or near zero and skewed the relative error.  \par

We note that calculating Bragg grating response with the ANN is much more computationally efficient than previously demonstrated methods.  This is because the Bragg ANN linearly increases in computation complexity with added grating parameters, while LDMTMM and all other methods known to these authors increase at least quadratically.    \par

To validate the Bragg ANN, we fabricated and measured several silicon photonic Bragg gratings with different chirping patterns and compared their transmission, reflection, and group delay spectra to the neural network's predictions. The gratings were arranged in one of two configurations: (1) a simple circuit that only measured the Bragg grating's transmission spectra and (2) a more complicated interrogation circuit capable of measuring the reflection, transmission, and group delay profiles from the same device simultaneously. Figure \ref{fig:panelFig_BG} illustrates the interrogation circuit used to measure all three responses simultaneously. In both configurations, grating couplers were used to route light on and off the chip. While the simpler circuit required less de-embedding, the full interrogator circuit allowed for a more comprehensive device characterization.

\begin{figure}
	\centering
	\includegraphics{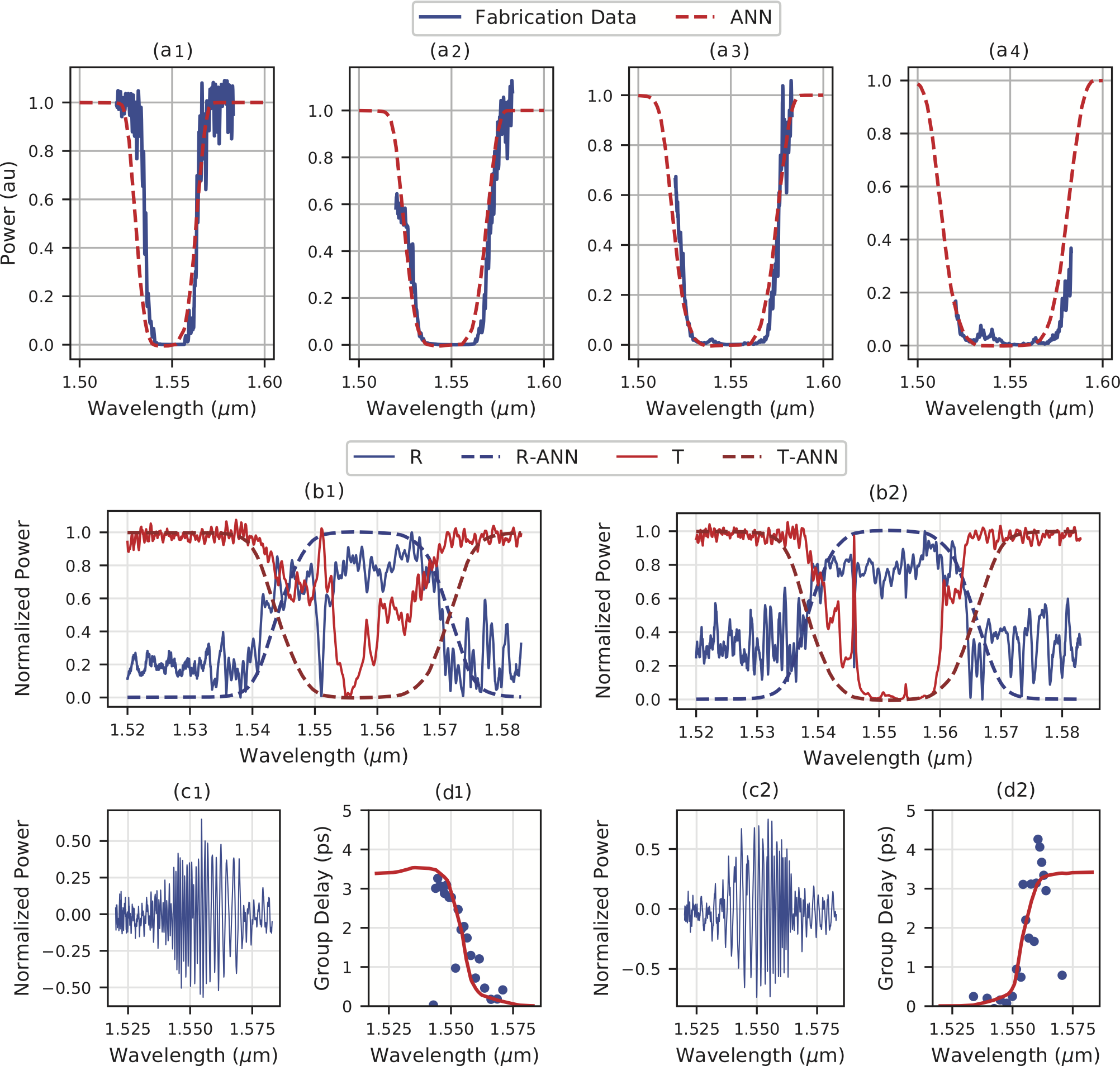}
	\caption{Fabrication data compared to corresponding ANN predictions. (a1-a4) Measured transmission responses for gratings with a period chirp of 5 nm (a1), 10 nm (a2), 15 nm (a3), and 20 nm (a4). (b1-b2) Transmission and reflection responses for two different Bragg gratings. Both gratings share the same design parameters, and have an identical but opposite linear chirp. The result of the mirrored chirping is seen in both the normalized MZI interference patterns (c1 , c2) and the extracted group delay responses (d1 , c2).}
	\label{fig:lukasData}
\end{figure}

The transmission-only gratings were designed with various grating period bandwidths from 5 nm to 20 nm, each with 600 periods and a corrugation width of 50 nm.  The initial design parameters produced ANN predictions that match the measured data extremely well. Small discrepancies in the grating responses are largely attributed to the grating's apodization profile and detector noise. Figure \ref{fig:lukasData} illustrates the comparison between the ANN's predictions and the measured data.

We designed the remaining gratings using a much smaller chirp bandwidth of 3 nm with 750 grating periods and a 30 nm corrugation width. We mirrored the orientation of half the gratings in order to measure both positive and negative sloped group delay profiles. Once measured, we normalized the data by de-embedding the responses from the various Y-branches, directional couplers, and grating couplers that complicate the measurement data. The process is explained in the Supplementary Material. Even with the rather complex transfer function, the transmission, reflection, and group delay profiles match the ANN's corresponding predictions well except for occasional resonant features caused by fabrication defects.  These defects are expected since the narrow bandwidth devices have a grating pitch with a fine discretization that approaches the e-beam raster grid resolution.   Small changes in grating pitch that don't align with the raster grid  occasionally produce weak Fabry-Perot resonance conditions visible in the data. These raster-induced defects also account for a small lateral shift (\textasciitilde1 nm) in the responses.  Even with these fabrication challenges, it is notable that the ANN successfuly predicts  the transmission, reflection, and group delay profiles simultaneously.  In fact, the ability to do so in noisy fabrication environments is one of the key advantages of the ANN and may allow for efficient parameter extraction where other methods fail.

\subsection*{Forward design}

\begin{figure}
    \centering
    \includegraphics{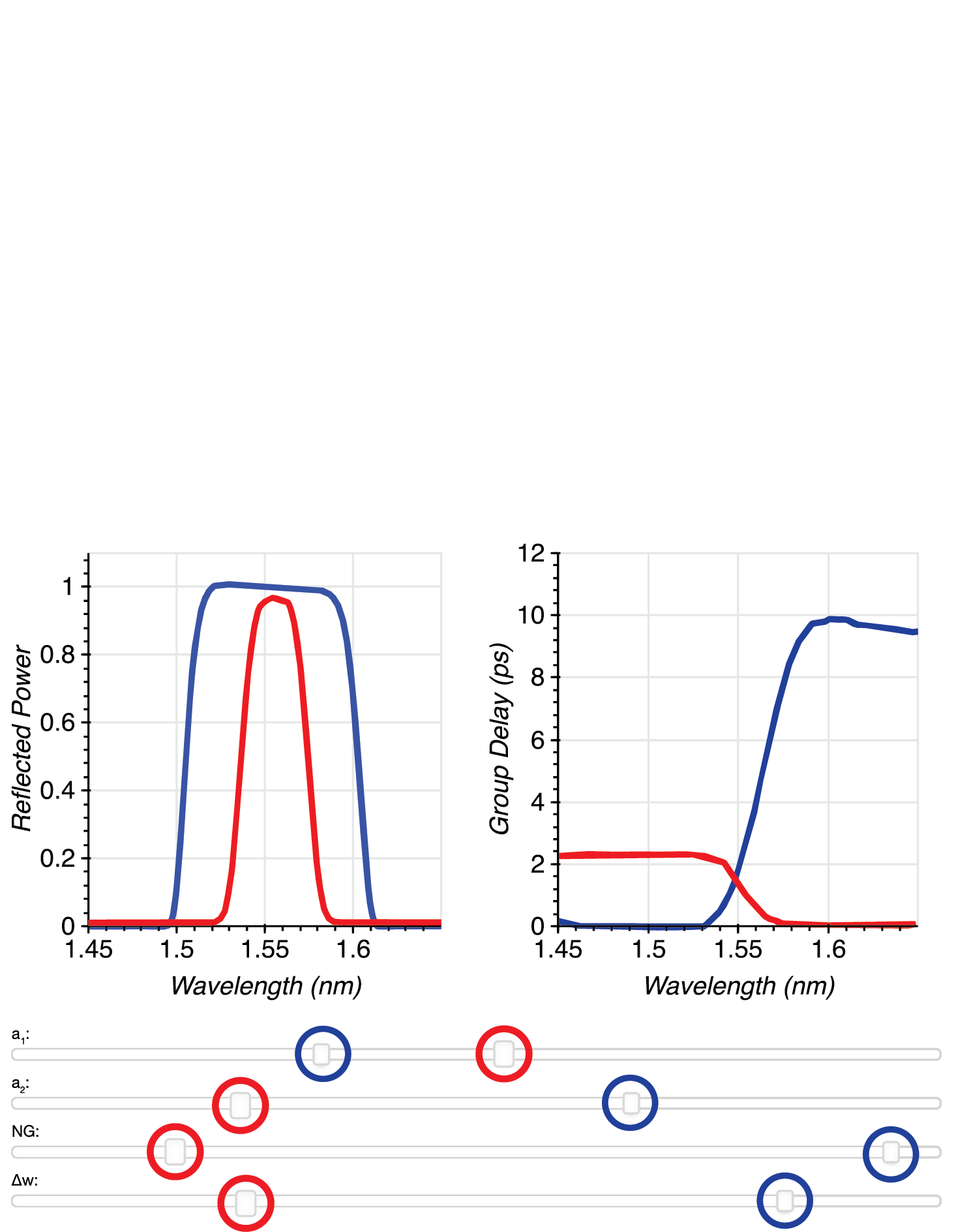}
    \caption{Graphical user interface used to explore the design space of a chirped Bragg grating. The slider bars on the left control physical parameters like grating length (NG), grating corrugation (dw), and the grating chirp (a1 and a2). Any time the user adjusts these parameters, the program calls the ANN and reproduces the expected reflection and group delay profiles for that particular grating. Due to the ANN's speed, the program is extremely responsive.}
    \label{fig:forward_design}
\end{figure}

The neural network's speed and flexibility enable forward design exploration. For example, Figure \ref{fig:forward_design} illustrates a graphical user interface (GUI) built with slider bars to adjust the Bragg grating's design parameters (i.e. corrugation widths, grating length, chirp pattern, etc). The plots dynamically update, calling the neural network every time the user modifies the input, and display the corresponding reflection and group delay profiles. Because wavelength is included as an input to the ANN rather than an output, arbitrary wavelength sampling within the domain is allowed. Computing these responses in real time is not possible using traditional techniques. This capability is valuable and allows even novice designers the ability to rapidly gain device intuition without necessarily understanding the underlying numerical techniques.

\subsection*{Inverse design}

\begin{figure}
    \centering
    \includegraphics{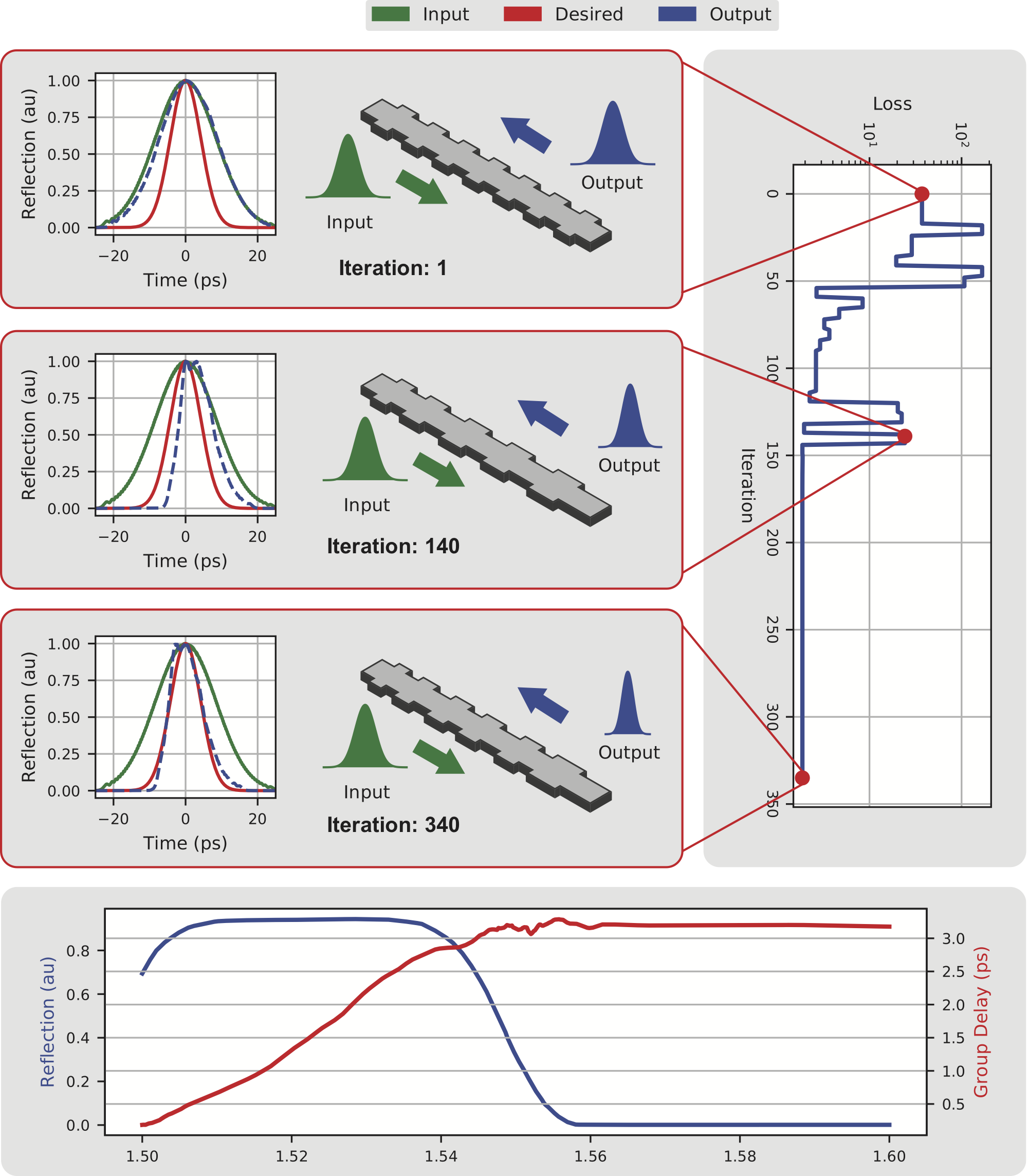}
    \caption{ANN-assisted design of a monolithic temporal pulse compressor using a silicon photonic chirped Bragg grating. A truncated Newton algorithm was tasked with constructing a grating that compressed an arbitrary chirped pulse by a factor of 2. After 340 grating simulations, the optimizer sufficiently minimized a cost function (right) that compared the new pulse's width to the old pulse. The resulting grating is demonstrated below and the input, output, and desired pulses for iterations 1, 140, and 340 are demonstrated on the left.}
    \label{fig:inverse_problem}
\end{figure}

This new approach also enables an entirely new set of inverse design problems. For example, we used the neural network in conjunction with a truncated Newtonian optimization algorithm to design a temporal pulse compressor. Designers often rely on dispersive Bragg gratings to generate short, optical pulses for high-capacity communications \cite{tan_monolithic_2010}. In this particular case study, we assumed an arbitrary source generates a 20 ps wide chirped pulse with 4 nm of bandwidth.  Figure \ref{fig:inverse_problem} illustrates the optimization routine's evolution, the resulting grating response, and the pulse shape before and after the Bragg grating. Such optimization algorithms run much quicker than previously known methods, owing to the accelerated cost function. The agnostic nature of the neural network interface works well with typical optimization routines, especially since any arbitrary wavelength sampling is allowed.  Depending on the cost function formulation, gradient-based methods could directly evaluate the Jacobian and Hessian tensors from the ANN without any extra sampling or discretization.

% ------------------------------------------------------- %
% Discussion - why we did what we did and why things happened
% ------------------------------------------------------- %

\section{Discussion}

Our method demonstrates a new, viable platform for silicon photonic circuit design. With a single global parameter fit, we successfully modeled silicon photonic waveguides and silicon photonic chirped Bragg gratings with arbitrary bandwidths, chirping patterns, lengths, and corrugation widths and allowed arbitrary wavelength sampling. Future work could explore new network architectures (e.g. different activation functions, layer connections, etc.) and training algorithms. Several other devices, like ring resonators \cite{bogaerts_silicon_2012}, can also be modeled. One could subsequently cascade several ANNs that model the scattering parameters of different devices, opening the door to large-scale optimization problems. \par

An important feature of this work is the choice to model the wavelength as a continuous input parameter rather than fix each output neuron at a specific wavelength point as done in all previous work known to the authors. The waveguide ANN, for example, outputs effective index values and the Bragg ANN outputs reflection and group delay values across the entire input spectrum. This approach, while more difficult to train, is more convenient for the designer. For example, an optimization routine tasked with designing a Bragg filter can focus more on parameters like the bandwidth and shape, rather than an arbitrarily sampled wavelength profile. Furthermore, this method doesn't require the training spectra to have the same sampling. Training sets for structures like ring resonators, whose features may require finer wavelength resolution than other devices, can now be strategically simulated to highlight these features. Assuming the network is trained correctly across a suitable domain, the ANN will seamlessly interpolate between both design parameters and wavelength points without any additional routines. \par

Unlike traditional simulation methods, training an arbitrary device ANN requires large datasets that are too time-intensive for most typical computers. With the growing availability of vast cloud-based computational resources,  however, several million training simulations can now be run in hours or days. \cite{vecchiola_high-performance_2009}.  Once trained, a neural network can reliably interpolate between training data, is compact and easily shared with the community, and can even continue to learn on new datasets via transfer learning \cite{pan_survey_2010}.  Thus, the computational complexity inherent in designing integrated photonic devices can be moved to the front end of the design process, allowing individual designers to work with abstracted components whose optical response can be rapidly calculated.

As with all deep learning applications, the network's utility is limited by biases introduced in the training set, the network architecture, or even the training process itself \cite{srivastava_dropout:_nodate}. Fortunately, we can anticipate these biases by extracting the model's prediction uncertainty without modifying our network architecture. Dropout inference techniques leverage models that rely on dropout layers to mitigate over-fitting (a form of network bias) \cite{gal_dropout_2015}. Even pre-trained networks can use dropout inference to extract prediction uncertainties without any modifications to the network. This particular network design methodology opens the door to many more applications, like training on fabricated device data. Foundry's that develop process design kits (PDKs), for example, can use this technique to model their fabrication processes while preserving their trade secrets.\par

% ------------------------------------------------------- %
% Methods - how we did what we did
% ------------------------------------------------------- %

\section{Methods}

\subsection*{Training data generation and preprocessing}

We generated our waveguide neural network's training set on a high performance computing cluster (HPC) using MEEP Photonic Bands (MPB) \cite{johnson_block-iterative_2001}, a finite difference eigenmode solver. The solver simulated 31 different waveguide widths from 350 nm to 1500 nm and 31 waveguide thicknesses ranging from 150 nm to 400 nm resulting in 961 different geometries. The solver simulated 200 distinct wavelength points in the range of 1400 nm to 1700 nm. The total number of training samples fed into the neural network was 192,200. 70\% of the data set was used as training samples and the remaining 30 \% was used as validation samples. Each sample had three inputs (width, thickness, and wavelength) and four corresponding outputs (effective indices for the first two TE and TM modes). No postprocessing was performed on the waveguide training data. \par

On the same HPC, we generated our Bragg training set by simulating 104,131 different gratings with the layered dielectric media transfer matrix method (LDMTMM) \cite{helan_comparison_2006}. The LDMTMM method models each individual section of the Bragg grating as an ideal waveguide and cascades each sections's corresponding transfer matrix to estimate the grating's response for each wavelength point of interest. We calculated each individual waveguide's effective index using the waveguide neural network. Our simulations swept through 10 different corrugation widths from 10 nm to 100 nm, 11 different grating lengths from 100 periods to 2000 periods, and 32 different chirping patterns. 

Once the grating spectra were generated, we fit the results to a generalized skewed Gaussian (see Supplementary Material) to reduce ringing and to generalize the grating's response to arbitrary apodization profiles. We found that without fitting, the resulting oscillations significantly complicate the training process and restrict the network's domain to a single apodization. We fit both the reflection spectrum and group delay responses to generalized Gaussians and resampled the results with 250 wavelength points from 1.45 $\mu$m to 1.65 $\mu$m. Since the nonlinear fitting routine occasionally failed, not all of the simulated gratings were appropriate for testing. After filtering through the results, we generated a database of 26,032,750 training samples.

\subsection*{Neural network design and training}

Both neural networks were trained on the same HPC cluster using Keras \cite{chollet_keras_2015} and Tensorflow \cite{martin_abadi_tensorflow:_2015}. Several hundred different architectures were tested. To gauge the effectiveness of each architecture, the mean-squared-error and coefficient of determination ($R^2$) metrics were used. The waveguide neural network that worked best had 4 hidden layers with 128 neurons, 64, neurons, 32 neurons, and 16 neurons. Each neuron used a hyperbolic tangent activation function. The Bragg grating neural network was designed with 10 hidden layers and 128 neurons each. RELU activation functions were used. Both networks were trained with 16 sample batch sizes. While the waveguide neural network was trained with 100 epochs, the Bragg grating neural network only needed about 5 epochs to reach sufficient results, primarily due to the large training set. The Bragg training set was normalized to improve the network's expressive capabilities.

\subsection*{Simulation benchmarks}
We performed all benchmarks using a quad-core Intel(R) i5-2400 CPU clocked at 3.10 GHz with 12 GB of RAM. To evaluate the waveguide ANN's speed, we simulated various waveguide parameters in serial using both the ANN and MPB. To evaluate the BG ANN's speed, we simulated various grating's in serial using both the ANN and the LDMTMM. We linearly fit each method's results and compared the slopes to examine the speedup.

\subsection*{Device fabrication}

The silicon photonic Bragg gratings were fabricated by Applied Nanotools Inc (Edmonton, Canada) using a direct-write 100 keV electron beam lithographic process. Silicon-on-Insulator wafers with 220 nm device thickness and 2 $\mu$m thick insulator layer were used. The devices were patterned with a raster step of 5 $\mu$m and etched with a ICP-RIE etch process. A 2.2 $\mu$m oxide cladding was deposited using a plasma-enhanced chemical vapour deposition (PECVD) process.

\subsection*{Device measurement}
Each device was measured using an automated process at the University of British Colombia (UBC). An Agilent 81600B tunable laser was used as the input source and Agilent 81635A optical power sensors as the output detectors. The wavelength was swept from 1500 to 1600 nm in 10 pm steps.  A polarization maintaining (PM) fibre was used to maintain the polarization state of the light, to couple the TE polarization in and out of the grating couplers. Several dembedded test structures were used to normalize out the coupler profiles.

\section*{Data Availability}
The data that support the plots within this paper and other findings of this study are available from the corresponding authors upon reasonable request.

\bibliographystyle{IEEEtran}
\bibliography{paper_ANN_RMC}

\begin{thebibliography}{10}
\providecommand{\url}[1]{#1}
\csname url@samestyle\endcsname
\providecommand{\newblock}{\relax}
\providecommand{\bibinfo}[2]{#2}
\providecommand{\BIBentrySTDinterwordspacing}{\spaceskip=0pt\relax}
\providecommand{\BIBentryALTinterwordstretchfactor}{4}
\providecommand{\BIBentryALTinterwordspacing}{\spaceskip=\fontdimen2\font plus
\BIBentryALTinterwordstretchfactor\fontdimen3\font minus
  \fontdimen4\font\relax}
\providecommand{\BIBforeignlanguage}[2]{{%
\expandafter\ifx\csname l@#1\endcsname\relax
\typeout{** WARNING: IEEEtran.bst: No hyphenation pattern has been}%
\typeout{** loaded for the language `#1'. Using the pattern for}%
\typeout{** the default language instead.}%
\else
\language=\csname l@#1\endcsname
\fi
#2}}
\providecommand{\BIBdecl}{\relax}
\BIBdecl

\bibitem{chrostowski_silicon_2015}
L.~Chrostowski and M.~Hochberg, \emph{\BIBforeignlanguage{English}{Silicon
  {Photonics} {Design}: {From} {Devices} to {Systems}}}, 1st~ed.\hskip 1em plus
  0.5em minus 0.4em\relax Cambridge ; New York: Cambridge University Press, May
  2015.

\bibitem{bogaerts_silicon_2018}
W.~Bogaerts and L.~Chrostowski, ``\BIBforeignlanguage{English}{Silicon
  {Photonics} {Circuit} {Design}: {Methods}, {Tools} and {Challenges}},''
  \emph{\BIBforeignlanguage{English}{Laser \& Photonics Reviews}}, vol.~12,
  no.~4, p. 1700237, Apr. 2018.

\bibitem{da_silva_ferreira_towards_2018}
A.~da~Silva~Ferreira, C.~H. da~Silva~Santos, M.~S. Gonçalves, and H.~E.
  Hernández~Figueroa, ``Towards an integrated evolutionary strategy and
  artificial neural network computational tool for designing photonic coupler
  devices,'' \emph{Applied Soft Computing}, vol.~65, pp. 1--11, Apr. 2018.

\bibitem{tahersima_deep_2018}
M.~H. Tahersima, K.~Kojima, T.~Koike-Akino, D.~Jha, B.~Wang, C.~Lin, and
  K.~Parsons, ``\BIBforeignlanguage{en}{Deep {Neural} {Network} {Inverse}
  {Design} of {Integrated} {Nanophotonic} {Devices}},''
  \emph{\BIBforeignlanguage{en}{arXiv:1809.03555 [physics.app-ph]}}, Sep. 2018.

\bibitem{inampudi_neural_2018}
S.~Inampudi and H.~Mosallaei, ``\BIBforeignlanguage{English}{Neural network
  based design of metagratings},'' \emph{\BIBforeignlanguage{English}{Applied
  Physics Letters}}, vol. 112, no.~24, p. 241102, Jun. 2018.

\bibitem{ferreira_computing_2018}
A.~D. Silva~Ferreira, G.~N. Malheiros-Silveira, and H.~E. Hernandez-Figueroa,
  ``\BIBforeignlanguage{English}{Computing {Optical} {Properties} of {Photonic}
  {Crystals} by {Using} {Multilayer} {Perceptron} and {Extreme} {Learning}
  {Machine}},'' \emph{\BIBforeignlanguage{English}{Journal of Lightwave
  Technology}}, vol.~36, no.~18, pp. 4066--4073, Sep. 2018.

\bibitem{zhang_spectrum_2018}
T.~Zhang, Q.~Liu, J.~Dai, X.~Han, J.~Li, Y.~Zhou, and K.~Xu, ``Spectrum
  prediction and inverse design for plasmonic waveguide system based on
  artificial neural networks,'' \emph{arXiv:1805.06410 [physics]}, May 2018.

\bibitem{peurifoy_nanophotonic_2018}
J.~Peurifoy, Y.~Shen, L.~Jing, Y.~Yang, F.~Cano-Renteria, B.~G. DeLacy, J.~D.
  Joannopoulos, M.~Tegmark, and M.~Soljacic,
  ``\BIBforeignlanguage{English}{Nanophotonic particle simulation and inverse
  design using artificial neural networks},''
  \emph{\BIBforeignlanguage{English}{Science Advances}}, vol.~4, no.~6, p.
  eaar4206, Jun. 2018.

\bibitem{rochette_gain_1999}
M.~Rochette, M.~Guy, S.~LaRochelle, J.~Lauzon, and F.~Trepanier,
  ``\BIBforeignlanguage{English}{Gain equalization of {EDFA}'s with {Bragg}
  gratings},'' \emph{\BIBforeignlanguage{English}{IEEE Photonics Technology
  Letters}}, vol.~11, no.~5, pp. 536--538, May 1999.

\bibitem{tan_chip-scale_2008}
D.~T.~H. Tan, K.~Ikeda, R.~E. Saperstein, B.~Slutsky, and Y.~Fainman,
  ``\BIBforeignlanguage{English}{Chip-scale dispersion engineering using
  chirped vertical gratings},'' \emph{\BIBforeignlanguage{English}{Optics
  Letters}}, vol.~33, no.~24, pp. 3013--3015, Dec. 2008.

\bibitem{strain_design_2010}
M.~J. Strain and M.~Sorel, ``\BIBforeignlanguage{English}{Design and
  {Fabrication} of {Integrated} {Chirped} {Bragg} {Gratings} for {On}-{Chip}
  {Dispersion} {Control}},'' \emph{\BIBforeignlanguage{English}{IEEE Journal of
  Quantum Electronics}}, vol.~46, no.~5, pp. 774--782, May 2010.

\bibitem{tan_monolithic_2010}
D.~T.~H. Tan, P.~C. Sun, and Y.~Fainman,
  ``\BIBforeignlanguage{English}{Monolithic nonlinear pulse compressor on a
  silicon chip},'' \emph{\BIBforeignlanguage{English}{Nature Communications}},
  vol.~1, p. 116, Nov. 2010.

\bibitem{b._j._eggleton_g._lenz_n._m._litc_optical_2000}
\BIBentryALTinterwordspacing
N.~M.~L. B.~J.~Eggleton, G.~Lenz, ``Optical {Pulse} {Compression} {Schemes}
  {That} {Use} {Nonlinear} {Bragg} {Gratings},'' \emph{Fiber and Integrated
  Optics}, vol.~19, no.~4, pp. 383--421, Oct. 2000. [Online]. Available:
  \url{https://doi.org/10.1080/014680300300001725}
\BIBentrySTDinterwordspacing

\bibitem{helan_comparison_2006}
R.~Helan, ``\BIBforeignlanguage{English}{Comparison of methods for fiber
  {Bragg} gratings simulation},'' ser. International {Spring} {Seminar} on
  {Electronics} {Technology} {ISSE}.\hskip 1em plus 0.5em minus 0.4em\relax
  IEEE, 2006, pp. 175+.

\bibitem{gallagher_eigenmode_2003}
D.~F.~G. Gallagher and T.~P. Felici, ``\BIBforeignlanguage{English}{Eigenmode
  expansion methods for simulation of optical propagation in photonics - {Pros}
  and cons},'' in \emph{\BIBforeignlanguage{English}{Integrated {Optics}:
  {Devices}, {Materials}, and {Technologies} {Vii}}}, Y.~S. Sidorin and
  A.~Tervonen, Eds.\hskip 1em plus 0.5em minus 0.4em\relax Bellingham: Spie-Int
  Soc Optical Engineering, 2003, vol. 4987, pp. 69--82.

\bibitem{bogaerts_silicon_2012}
W.~Bogaerts, P.~De~Heyn, T.~Van~Vaerenbergh, K.~De~Vos, S.~K. Selvaraja,
  T.~Claes, P.~Dumon, P.~Bienstman, D.~Van~Thourhout, and R.~Baets,
  ``\BIBforeignlanguage{English}{Silicon microring resonators},''
  \emph{\BIBforeignlanguage{English}{Laser \& Photonics Reviews}}, vol.~6,
  no.~1, pp. 47--73, Jan. 2012.

\bibitem{vecchiola_high-performance_2009}
C.~Vecchiola, S.~Pandey, and R.~Buyya,
  \emph{\BIBforeignlanguage{English}{High-{Performance} {Cloud} {Computing}:
  {A} {View} of {Scientific} {Applications}}}.\hskip 1em plus 0.5em minus
  0.4em\relax New York: IEEE, 2009.

\bibitem{pan_survey_2010}
S.~J. Pan and Q.~Yang, ``A {Survey} on {Transfer} {Learning},'' \emph{IEEE
  Transactions on Knowledge and Data Engineering}, vol.~22, no.~10, pp.
  1345--1359, Oct. 2010.

\bibitem{srivastava_dropout:_nodate}
N.~Srivastava, G.~Hinton, A.~Krizhevsky, I.~Sutskever, and R.~Salakhutdinov,
  ``\BIBforeignlanguage{English}{Dropout: {A} {Simple} {Way} to {Prevent}
  {Neural} {Networks} from {Overfitting}},''
  \emph{\BIBforeignlanguage{English}{Journal of Machine Learning Research}},
  vol.~15, pp. 1929--1958, Jun. 2014.

\bibitem{gal_dropout_2015}
Y.~Gal and Z.~Ghahramani, ``Dropout as a {Bayesian} {Approximation}:
  {Representing} {Model} {Uncertainty} in {Deep} {Learning},''
  \emph{arXiv:1506.02142 [cs, stat]}, Jun. 2015.

\bibitem{johnson_block-iterative_2001}
S.~G. Johnson and J.~D. Joannopoulos,
  ``\BIBforeignlanguage{English}{Block-iterative frequency-domain methods for
  {Maxwell}'s equations in a planewave basis},''
  \emph{\BIBforeignlanguage{English}{Optics Express}}, vol.~8, no.~3, pp.
  173--190, Jan. 2001.

\bibitem{chollet_keras_2015}
\BIBentryALTinterwordspacing
``Keras.'' [Online]. Available: \url{https://keras.io/}
\BIBentrySTDinterwordspacing

\bibitem{martin_abadi_tensorflow:_2015}
\BIBentryALTinterwordspacing
{TensorFlow}: {Large}-{Scale} {Machine} {Learning} on {Heterogeneous}
  {Systems}. [Online]. Available: \url{http://tensorflow.org/}
\BIBentrySTDinterwordspacing

\end{thebibliography}


\begin{thebibliography}{1}
\providecommand{\url}[1]{#1}
\csname url@samestyle\endcsname
\providecommand{\newblock}{\relax}
\providecommand{\bibinfo}[2]{#2}
\providecommand{\BIBentrySTDinterwordspacing}{\spaceskip=0pt\relax}
\providecommand{\BIBentryALTinterwordstretchfactor}{4}
\providecommand{\BIBentryALTinterwordspacing}{\spaceskip=\fontdimen2\font plus
\BIBentryALTinterwordstretchfactor\fontdimen3\font minus
  \fontdimen4\font\relax}
\providecommand{\BIBforeignlanguage}[2]{{%
\expandafter\ifx\csname l@#1\endcsname\relax
\typeout{** WARNING: IEEEtran.bst: No hyphenation pattern has been}%
\typeout{** loaded for the language `#1'. Using the pattern for}%
\typeout{** the default language instead.}%
\else
\language=\csname l@#1\endcsname
\fi
#2}}
\providecommand{\BIBdecl}{\relax}
\BIBdecl

\bibitem{schmidhuber_deep_2015}
J.~Schmidhuber, ``\BIBforeignlanguage{English}{Deep learning in neural
  networks: {An} overview},'' \emph{\BIBforeignlanguage{English}{Neural
  Networks}}, vol.~61, pp. 85--117, Jan. 2015.

\bibitem{cun_theoretical_1988}
Y.~le~Cun, ``\BIBforeignlanguage{English (US)}{A theoretical framework for
  back-propagation},'' \emph{\BIBforeignlanguage{English (US)}{Proceedings of
  the 1988 Connectionist Models Summer School, CMU, Pittsburg, PA}}, pp.
  21--28, 1988.

\bibitem{hornik_approximation_1991}
K.~Hornik, ``\BIBforeignlanguage{English}{Approximation {Capabilities} of
  {Multilayer} {Feedforward} {Networks}},''
  \emph{\BIBforeignlanguage{English}{Neural Networks}}, vol.~4, no.~2, pp.
  251--257, 1991.

\bibitem{hill_fiber_1997}
K.~O. Hill and G.~Meltz, ``Fiber {Bragg} grating technology fundamentals and
  overview,'' \emph{Journal of Lightwave Technology}, vol.~15, no.~8, pp.
  1263--1276, Aug. 1997.

\bibitem{harris_use_1978}
F.~J. Harris, ``On the use of windows for harmonic analysis with the discrete
  {Fourier} transform,'' \emph{Proceedings of the IEEE}, vol.~66, no.~1, pp.
  51--83, Jan. 1978.

\bibitem{Storn1997}
\BIBentryALTinterwordspacing
R.~Storn and K.~Price, ``Differential evolution -- a simple and efficient
  heuristic for global optimization over continuous spaces,'' \emph{Journal of
  Global Optimization}, vol.~11, no.~4, pp. 341--359, Dec 1997. [Online].
  Available: \url{https://doi.org/10.1023/A:1008202821328}
\BIBentrySTDinterwordspacing

\bibitem{marquardt_algorithm_1963}
\BIBentryALTinterwordspacing
D.~Marquardt, ``An {Algorithm} for {Least}-{Squares} {Estimation} of
  {Nonlinear} {Parameters},'' \emph{Journal of the Society for Industrial and
  Applied Mathematics}, vol.~11, no.~2, pp. 431--441, Jun. 1963. [Online].
  Available: \url{https://epubs.siam.org/doi/10.1137/0111030}
\BIBentrySTDinterwordspacing

\bibitem{chrostowski_silicon_2015}
L.~Chrostowski and M.~Hochberg, \emph{\BIBforeignlanguage{English}{Silicon
  {Photonics} {Design}: {From} {Devices} to {Systems}}}, 1st~ed.\hskip 1em plus
  0.5em minus 0.4em\relax Cambridge ; New York: Cambridge University Press, May
  2015.

\end{thebibliography}

\section*{Acknowledgments}
The authors would like to thank Lukas Chrostowski for useful discussions relating to the Bragg structures and for facilitating the SiEPIC fabricating process, as well as David Buck for supplying additional fabrication data.

\section*{Author Contributions}
AMH \& RMC conceived the idea, AMH designed and trained the ANN, AMH \& RMC designed the devices, AMH \& RMC evaluated and interpreted the data, AMH \& RMC wrote the manuscript.

\section*{Competing Interests}
The authors declare no competing interests.

\section*{Materials \& Correspondence}
Please send all correspondence and data requests to camacho@byu.edu

\end{document}

\maketitle

\section*{Training Data Processing}
Artificial neural networks model the relationship between their inputs and outputs by cascading various computational units, known as neurons \cite{schmidhuber_deep_2015}.  ANNs learn the desired relationship between inputs and outputs by tuning each neuron in response to a training set.  Training algorithms, such as backpropagation, are used to strategically introduce these datasets into the neural network and gradually update the network's parameters until a convergence criteria is met \cite{cun_theoretical_1988}. If the network architecture is sufficiently large, any arbitrary relationship manifested in the dataset can be modeled \cite{hornik_approximation_1991}. Capturing complex behavior, like parameterized sinusoidal ringing, requires sufficient sampling.

\begin{figure}
	\centering
	\includegraphics{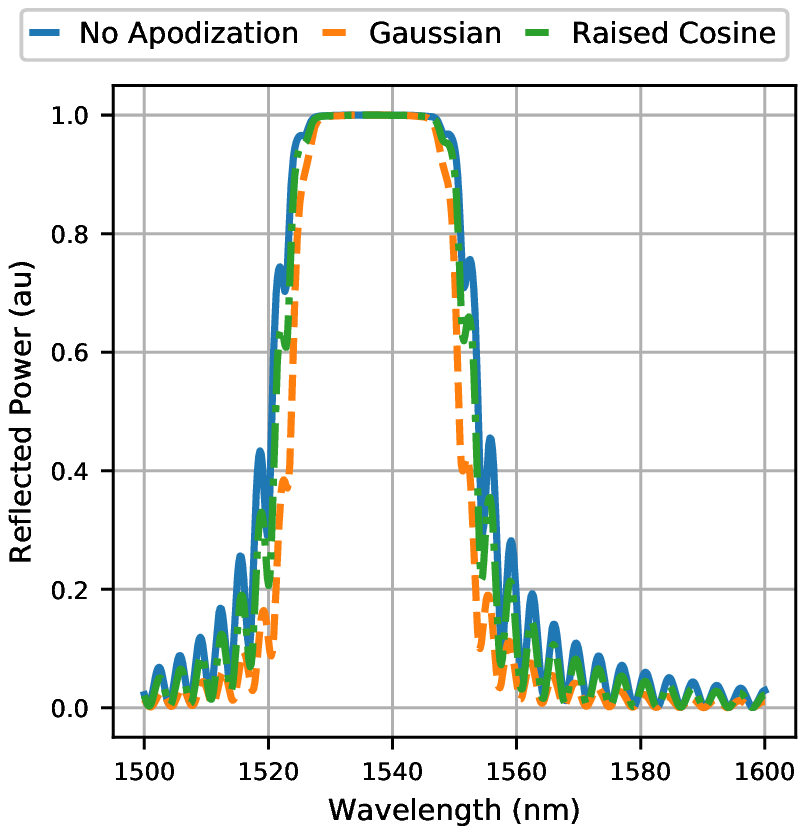}
	\caption{Reflection profile for an integrated chirped Bragg grating with no apodization (blue), a Gaussian apodization (orange), and a raised cosine apodization. Different apodization functions reduce the response's ringing.}
	\label{fig:apodization}
\end{figure}

Bragg grating device responses are heavily influenced by their apodization \cite{hill_fiber_1997}. For example, devices without apodization have much more ringing than devices with a raised cosine apodization. The effect is similar to finite impulse response (FIR) filtering windows \cite{harris_use_1978}. Figure \ref{fig:apodization} illustrates various apodization windows and the resultant ringing. While the layered dielectric medium transfer matrix method (LDMTMM) can effectively simulate such ringing, fabrication inconsistencies make predicting the actual ringing close to impossible. \par

Consequently, designers focus more effort on shaping the Bragg grating's reflection, transmission, and group delay responses than mitigating ringing. This mentality encourages a design tool that temporarily ignores any ringing and simulates the basic grating response profiles. \par

\begin{figure}
	\centering
	\includegraphics[width=6in]{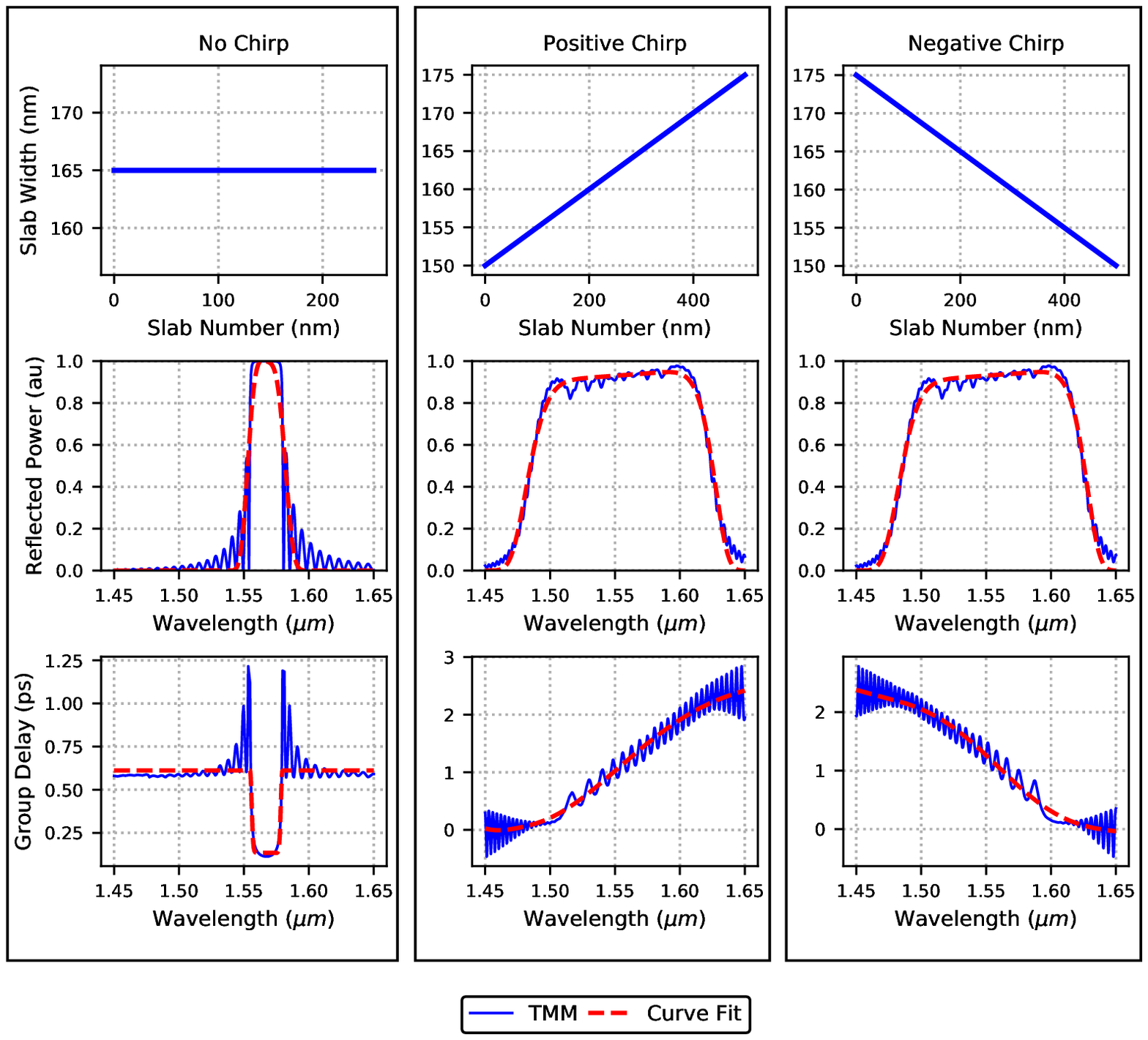}
	\caption{Demonstration of the fitting algorithm for a Bragg grating with no chirp (column one), a postive chirp (column two) and a negative chirp (column three). The chirp patterns themselves are illustrated in the first row, the reflection profiles are depicted in the second row, and the respective group delay responses are found in the third row. The modified Gaussian function accounts for the wider bandwidths, skewness, and overall shape of the responses will ``filter" out the apodization dependent ringing.}
	\label{fig:fitFigure}
\end{figure}

To accomplish this, we filtered all apodization dependent ringing from the Bragg ANN's training set by fitting the reflection spectra and group delay responses. Through trial and error, we found that a generalized, skewed Gaussian of the form
\begin{equation}
	f(\lambda,\lambda_0,\sigma,\beta,a,p,c) = \frac{a \sigma}{\gamma} e ^{\frac{-\beta|\lambda - \lambda_0|}{\gamma} ^ p} + c
\end{equation}
where 
\begin{equation}
	\gamma = \frac{2\sigma}{1+e^{-\beta (\lambda-\lambda_0)}}
\end{equation}
most comprehensively models both the reflection and group delay profiles for a wide array of tuning parameters. Each parameter strategically tunes a particular feature commonly found in both responses. For example, $a$ simply scales the maximum reflectivity or maximum group delay, $\sigma$ shapes the reflection bandwidth or group delay spread, $\beta$ induces skewness to one side or the other, $p$ flattens the function's main lobe (i.e. for saturated reflection profiles), and $c$ provides the necessary offsets for the group delay profiles. Figure \ref{fig:fitFigure} illustrates various LDMTMM simulations and the corresponding fit. \par

To find the optimal coefficients, we first ran a differential evolution algorithm that minimized the mean squared error between the LDMTMM simulation and the functional fit \cite{Storn1997}. This global optimization routine provided a suitable starting point for a Levenberg-Marquardt algorithm to locate the local minima \cite{marquardt_algorithm_1963}. Using a two-staged optimization approach increased the robustness of the training processing and enabled batched parallelization on the high performance cluster (HPC). \par

\section*{Measurement Data Normalization}
The circuit used to simultaneously extract the transmission, reflection, and group delay profiles of the integrated Bragg gratings incorporated several additional devices that skewed the Bragg grating's transfer function. The grating couplers, Y-branches, and directional couplers used to route the signal all have non uniform frequency responses, and must be normalized out in order to accurately measure the Bragg grating's response. \par

Many designers fabricate de-embedded devices where each individual transfer function can be extracted and subsequently eliminated from the larger circuit \cite{chrostowski_silicon_2015}. Fabrication variability, however, prevents consistency from one device to another. The circuit's grating couplers, for example, sometimes showed 3 dB of variation across the band from one de-embedded structure to another. Consequently, this method cannot be reliably used to calibrate the Bragg grating responses. \par

\begin{figure}
	\centering
	\includegraphics{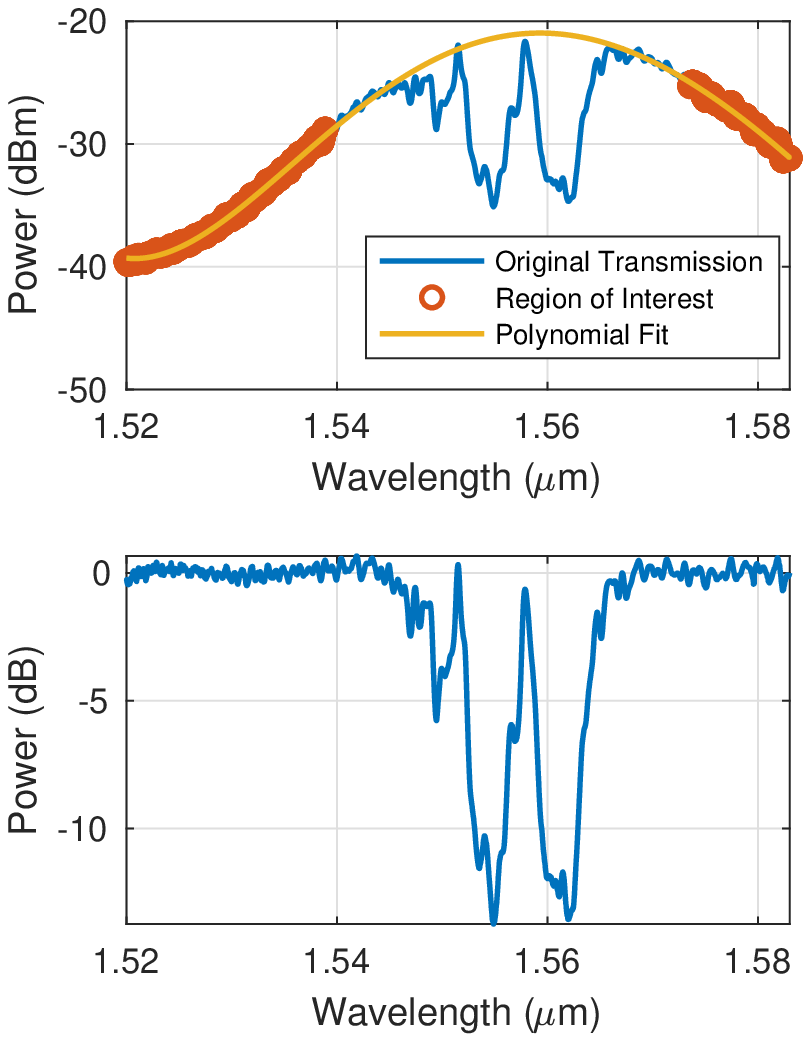}
	\caption{The normalization process for the integrated Bragg gratings. Data points outside of the stop band are fitted to a fifth degree polynomial to capture the response of the grating couplers and other devices within the circuit (top). The polynomial is then normalized from the data (bottom).}
	\label{fig:normalization}
\end{figure}

Instead, we fit the signal outside of the Bragg grating's reflection/transmission band using a fifth order polynomial that much more accurately describes the other devices' spectral influences. Figure \ref{fig:normalization} illustrates this procedure. Since the Bragg grating's group delay and magnitude responses are band-limited, we know what the rest of the spectrum should look like.

\begin{figure}
	\centering
	\includegraphics{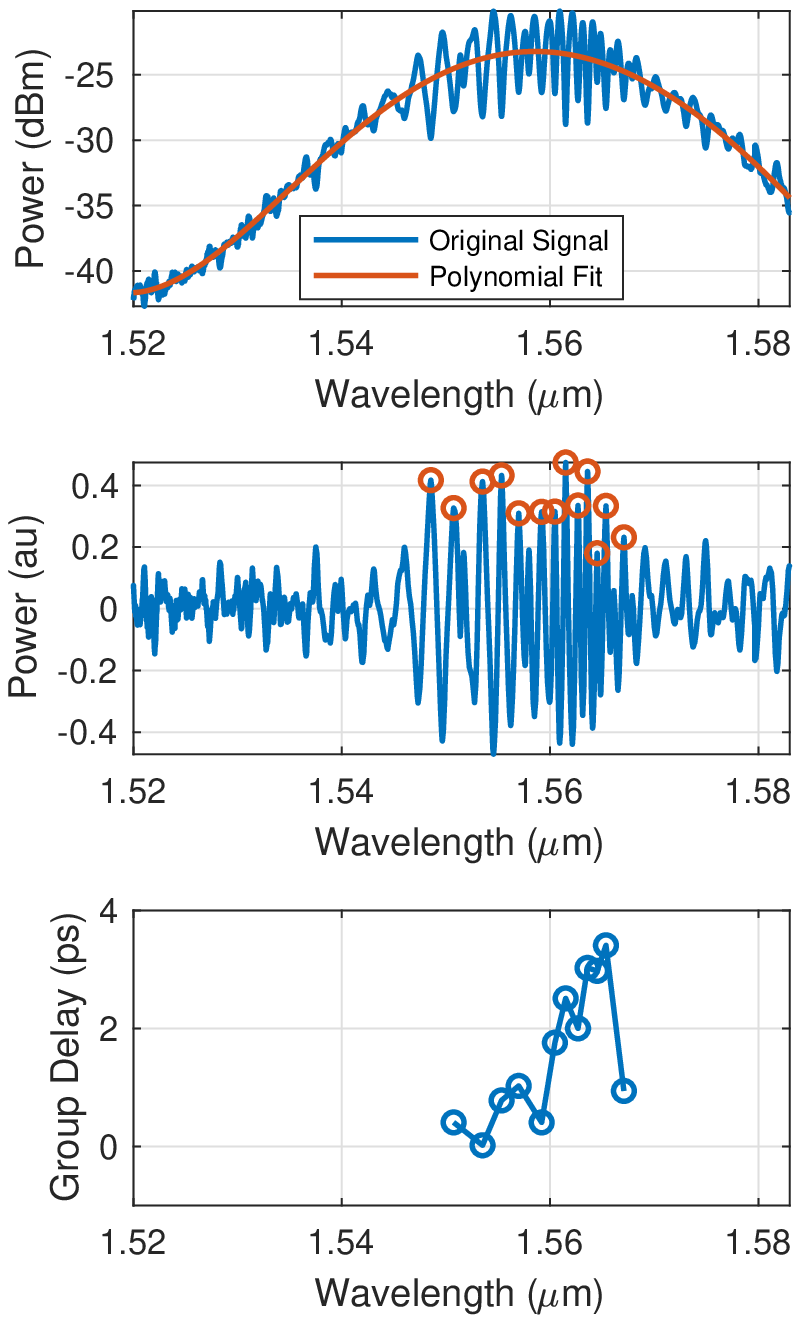}
	\caption{The group delay extraction process. First, the low frequency carrier is removed by using a fifth order polynomial fit (top). Next, the oscillation peaks are tracked and the FSR is estimated (middle). Finally, the group delay is calculated using the transfer function of the MZI (bottom).}
	\label{fig:groupDelay}
\end{figure}

To extract the group delay from the interference pattern, we first normalize out the low frequency carrier by fitting the entire signal to a fifth degree polynomial. Then, we estimate the free spectral range (FSR) by tracking each oscillation peak. Using the transfer function of the MZI \cite{chrostowski_silicon_2015} along with the FSR, the resulting group index and group delay is estimated. Figure \ref{fig:groupDelay} illustrates this process.

\bibliographystyle{IEEEtran}
\bibliography{supp_material}